\documentclass[12pt]{article}
\usepackage{times}
\usepackage{geometry}
\usepackage[utf8]{inputenc}
\usepackage{enumitem,amssymb}
\usepackage{ragged2e}
\usepackage{hyperref}
\usepackage{natbib}
\usepackage{multicol}
\usepackage{graphicx}
\usepackage{xcolor}
\usepackage{floatrow}
\newlist{thematic}{itemize}{8}
\setlist[thematic]{label=$\square$}
\usepackage{pifont}
\def\aj{AJ}%
\def\araa{ARA\&A}%
\def\apj{ApJ}%
\def\apjl{ApJ}%
\def\apjs{ApJS}%
\def\aap{A\&A}%
\def\mnras{MNRAS}%
\def\pasj{PASJ}%
\def\ssr{Space~Sci.~Rev.}%
\def\nat{Nature}%
\begin{document}
\huge
\begin{center}
Astro2020 Science White Paper: \\
A Shocking Shift in Paradigm \\
for Classical Novae
\end{center}
\normalsize

\noindent \textbf{Thematic Areas:} \hspace*{60pt} $\square$ Planetary Systems \hspace*{10pt} $\square$ Star and Planet Formation \hspace*{20pt}\linebreak
$\boxtimes$ Formation and Evolution of Compact Objects \hspace*{31pt} $\square$ Cosmology and Fundamental Physics \linebreak  $\square$ Stars and Stellar Evolution \hspace*{1pt} $\square$ Resolved Stellar Populations and their Environments \hspace*{40pt} \linebreak
$\square$    Galaxy Evolution   \hspace*{45pt}  $\boxtimes$ Multi-Messenger Astronomy and Astrophysics \hspace*{65pt} \linebreak
  
\textbf{Principal Author:}\\
Name:	Laura Chomiuk\\
Institution:  
Michigan State University\\
Email: chomiuk@pa.msu.edu\\
Phone: +1 (517) 884-5608 \\

\textbf{Co-authors:}\\
Elias Aydi (Michigan State University)\\
Aliya-Nur Babul (Columbia University)\\
Andrea Derdzinski (Columbia University) \\
Adam Kawash (Michigan State University)\\
Kwan-Lok Li (Ulsan National Institute of Science and Technology)\\
Justin Linford (West Virginia University)\\
Brian D.~Metzger (Columbia University)\\
Koji Mukai (NASA/GSFC and UMBC)\\
Michael P. Rupen (Herzberg Astronomy and Astrophysics, National Research Council Canada) \\
Jennifer Sokoloski (LSST Corporation \& Columbia University)\\
Kirill Sokolovsky (Michigan State University)\\
Elad Steinberg (Columbia University)\\

\textbf{Abstract}\\
The discovery of GeV $\gamma$-rays from classical novae has led to a reassessment of these garden-variety explosions, and highlighted their importance for understanding radiative shocks, particle acceleration, and dust formation in more exotic, distant transients. Recent collaboration between observers and theorists has revealed that shocks in novae are energetically important, and can even dominate their bolometric luminosity. Shocks may also explain long-standing mysteries in novae such as dust production, super-Eddington luminosities, and ``flares'' in optical light curves. Here, we highlight the multi-wavelength facilities of the next decade that will further test our nova shock model and fulfill the promise of novae as powerful astrophysical laboratories.

\pagebreak

\noindent{\large \bf 1. Novel Developments in Nova Science}\\
When the previous Decadal Survey was prepared ten years ago, study of classical novae seemed like a dying field of research. Owing to the apparent lack of progress in previous years, classical novae were not mentioned in the ``New Worlds, New Horizons'' report, unlike their more energetic cousins -- supernovae. This perception changed abruptly when GeV $\gamma$-ray emission was detected from the Galactic nova V407\,Cyg, 
making it clear that novae can be efficient accelerators of relativistic particles, and highlighting how much we still have to learn about novae \citep{Abdo10}.

\begin{figure}[b]
  \floatbox[{\capbeside\thisfloatsetup{capbesideposition={right,top},capbesidewidth=8.0cm}}]{figure}[\FBwidth]
  {\caption{\footnotesize \textbf{Nova shocks are orders of magnitude denser than other shocks in our Galactic neighborhood, and probe physical conditions present in intriguing classes of transients.} Shock parameters of novae are compared with the interplanetary medium on AU scales (IPM), the intergalactic medium (IGM), supernova remnants, radio-emitting blast waves of core-collapse SNe, gamma-ray burst afterglows (GRB), shock-powered supernovae (Type IIn and Type Ia-CSM), luminous red novae (non-degenerate stellar mergers), white dwarf--white dwarf mergers, and shock break-out from red supergiant (RSG) and blue supergiant stars (BSG). Figure adapted from \citet{Metzger16}.}  \label{shockparam}}
 {\includegraphics[width=8.0cm,angle=0]{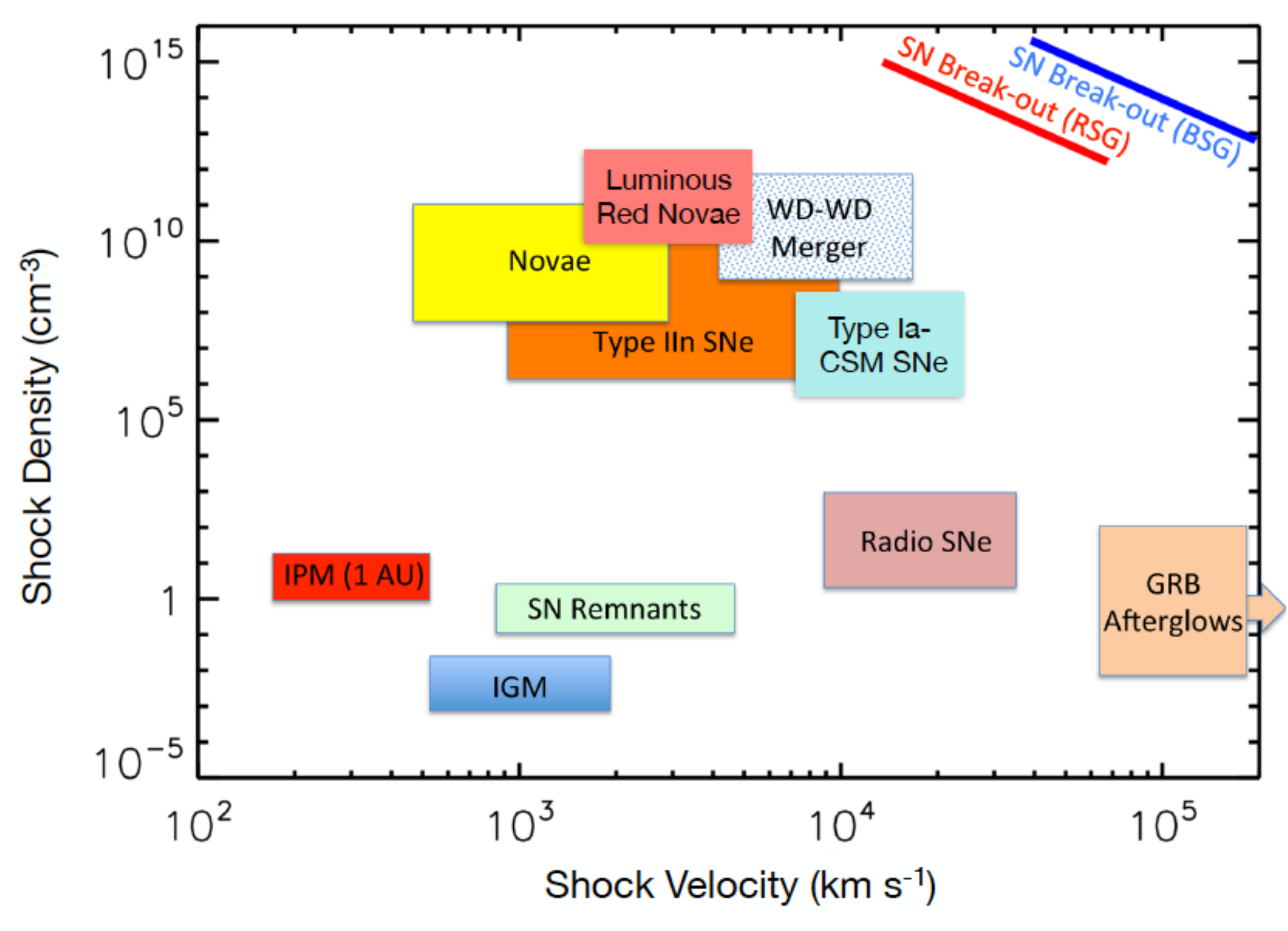}} 
\end{figure}

Since then, a paradigm shift has occurred in our understanding of novae: we are now finding that their luminosities (at least in some cases) are dominated by shock interactions, rather than thermal radiation from the nuclear-burning white dwarf \citep{Li17}. Meanwhile, the shocks in novae have been shown to be internal to the ejecta, and occurring at very high densities ($\gtrsim 10^{9}$ cm$^{-3}$; orders of magnitude higher than 
the ones commonly found in shocked interstellar or circumstellar material; Figure~\ref{shockparam}). This means that novae are important laboratories for understanding the optically-thick, radiative shocks that are critical in dictating the characteristics of more distant, exotic explosions like stellar mergers and super-luminous supernovae \citep{Metzger16}.

To answer fundamental questions about novae and fulfill their promise as  laboratories for understanding more exotic transients, this new paradigm  needs to be further tested and quantified with multi-wavelength observations in the next decade. In the following we briefly review our current understanding of nova physics, describe outstanding open questions in the field, and summarize future capabilities that will help us answer these questions.

\vspace{0.15in}
\noindent {\large \bf 2. Nova Science: Primed for the Next Decade of Discovery}\\
Nova explosions occur on accreting white dwarf in binary systems, powered by nuclear fusion that ignites at the bottom of the accreted hydrogen-rich shell \citep{2008clno.book.....B}.  The thermonuclear runaway leads to ejection of most (if not all) of this  $10^{-7}$--$10^{-3}$ M$_{\odot}$  shell at velocities of $\sim$500-5,000\,km/s.
After explosion, the binary remains intact and accretion is reinstated, implying that all novae recur---but on widely varying timescales of $1-10^8$ yr \citep{Yaron05}. 

Nuclear burning is sustained on the white dwarf surface for weeks to years following the thermonuclear runaway, serving as a near-Eddington luminosity radiation source ($\sim 10^{38}$ erg/s) in the center of the expanding ejecta \citep{Wolf13}. The paradigm until recently was that novae are essentially constant-luminosity events, and that the evolution of a nova outburst was due to the expansion of the ejecta \citep{Gallagher78}. Early in the outburst, the ejecta are optically thick, and this luminosity predominantly emerges in the optical; as the ejecta thin, the peak of the spectrum shifts to the ultraviolet, and finally to the soft X-rays, when we can directly see the white dwarf's photosphere \citep[e.g.,][]{Schwarz11}.


However, there have long been poorly-understood observational complications to this picture. Many novae do not appear as single ejections, but instead show multiple peaks in their optical light curves and multiple components in optical spectra, implying that several discrete ejections take place in the weeks following thermonuclear runaway \citep[e.g.,][]{Pejcha09, Tanaka11}. X-ray observations of novae reveal the presence of hot 
($\gtrsim 10^7$~K) 
gas, indicating the presence of shocks \citep[e.g.][]{1994MNRAS.271..155O,2001ApJ...551.1024M,Nelson19}.
However, the prevalence, strength, and energetic importance of  these shocks were not appreciated until \emph{Fermi}/LAT began routinely detecting GeV $\gamma$-rays from Galactic novae.

Since 2010, \emph{Fermi} has detected 14 novae emitting GeV $\gamma$-rays, with a few weeks around optical maximum \citep{Abdo10, Ackermann14, Franckowiak18}. While a couple of these systems have red giant companions, most have dwarf companions and are surrounded by very low-density circumstellar material. This implies that the shocks that accelerate particles to relativistic speeds and produce $\gamma$-rays are \emph{internal} to the ejecta. Further support for this picture comes from radio imaging of the thermal ejecta and synchrotron-emitting shocked regions \citep[][\S3]{Chomiuk14}, and the very large columns of material that are observed to absorb the shock's X-ray emission \citep[][\S5]{Nelson19}.

The shocks occur early in the outburst, in ejecta expanding at modest velocities---implying that the densities are high (Figure 1). \citet{Metzger14} predicted that the densities may be so high that the shocks are radiative: the shocked gas may not primarily emit in the X-ray, but instead dwell at $\sim 10^4$ K and radiate in the optical (see also \citealt{Steinberg18}). This prediction was borne out when \cite{Li17} found correlated variations in the $\gamma$-ray and optical bands for the nova V5856\,Sgr, suggesting that the optical emission, like the $\gamma$-rays, is originating in shocks. Because the nova spectral energy distribution peaks in the optical at these early times, the implication is that a significant fraction of the nova's bolometric luminosity came from reprocessed energy of shocks. Another intriguing theoretical prediction is that these dense radiative shocks may be the locations of often-observed dust formation in novae---a decades old unsolved problem, given the warm, expanding, irradiated environments of nova ejecta \citep[][\S6]{Derdzinski17}.

We have only just begun to understand the cause, energetics, and implications of shocks in novae, and many critical questions remain to be solved in the next decade.

\vspace{0.15in}
\noindent {\large \bf 3. Where and Why do Nova Shocks Form?}\\
The physical drivers of prolonged, complex mass loss in novae has long been a mystery, founded in some of the most difficult and important problems that plague stellar astrophysics today---like common envelope ejection and wind mass loss prescriptions. The need to understand the physics driving nova mass loss has a renewed urgency, with the realization that nova shocks are powered by this complex mass ejection and are energetically important. 

We have developed a hypothesis for where shocks form in the nova ejecta, which needs to be tested and honed in the next decade. Based on radio imaging of V959~Mon (2012), our picture is that binary interaction with the nova envelope produces a relatively slowly expanding equatorial torus. Later, the white dwarf powers a fast bipolar wind. Shocks and $\gamma$-rays are produced at the interface of these two outflows \citep[Figure 2,][]{Chomiuk14}.  The Next Generation Very Large Array (ngVLA; \citealt{Murphy18}) will be the ideal instrument for imaging shocks in novae. With its order-of-magnitude improvement in sensitivity over the current VLA, and a diversity of baseline lengths extending to 1000 km, every ngVLA observation of a nova will have the potential to reveal structure like that imaged in Figure 2--- simultaneously tracing both thermal emission from the warm $\sim 10^{4}$~K ejecta and synchrotron emission from the shocked regions. 

 \begin{figure}[t]\centering
 \includegraphics[width=13cm]{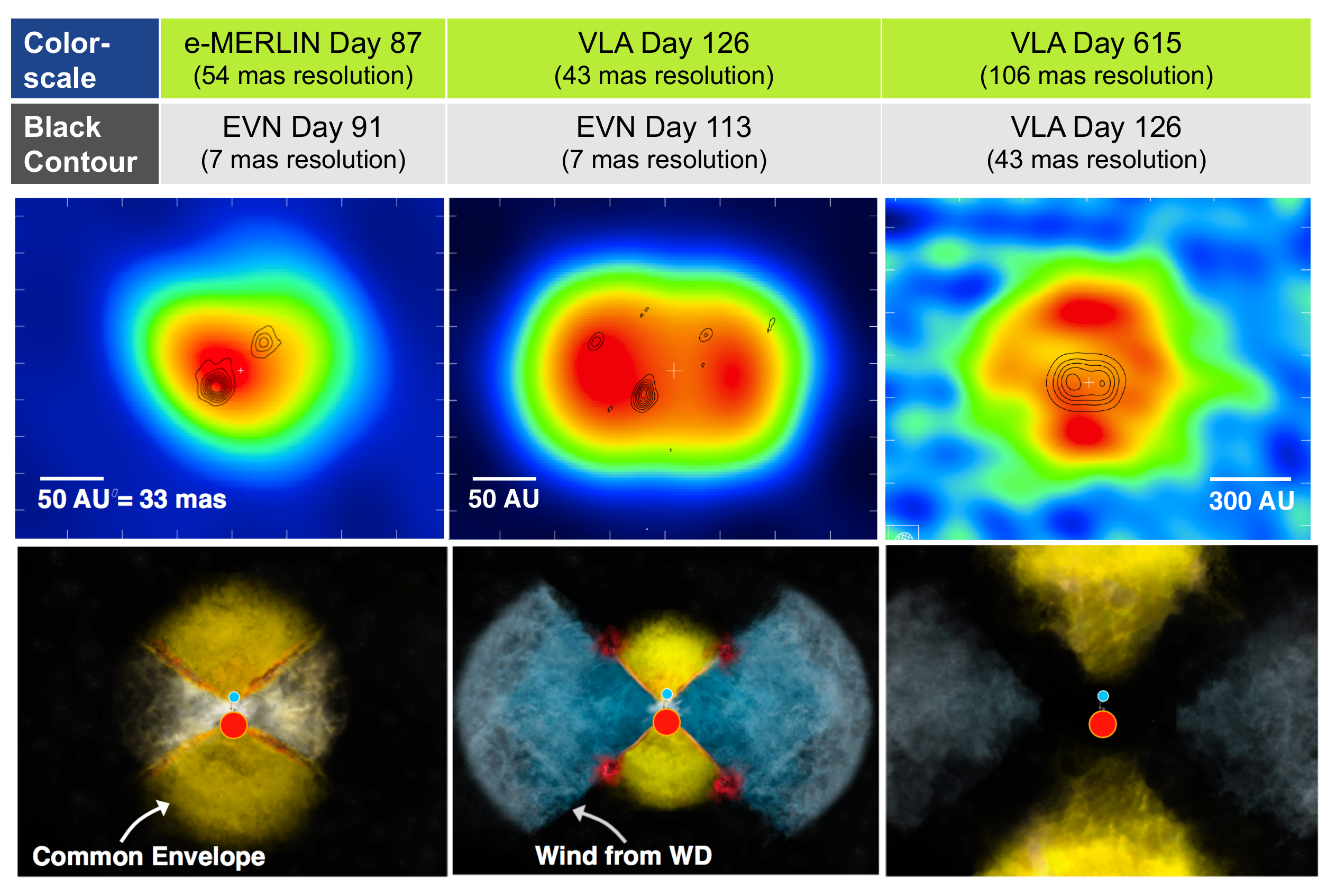}
 \vspace{-0.20cm}
 \caption{\footnotesize \textbf{Images of nova V959~Mon show ejecta geometry consistent with models of common envelope evolution and shocks from colliding flows.} The top row are radio observations (\citealt{Chomiuk14}; labelled at top), and the bottom row is a series of artist's representations (B.\ Saxton/NRAO/AUI/NSF). {\bf Left panels:} the thermal nova ejecta are barely resolved at early times (colorscale in top row), while synchrotron knots are superimposed on the ejecta (black contours in top row).  {\bf Middle:} It is now clear that the synchrotron knots surround two thermal lobes. We interpret this as  a fast wind which has been funneled along the binary's poles, crashing into the equatorial material and producing shocks and $\gamma$-ray emission at the interface (red regions in artist's representation). {\bf Right:} The polar outflow has faded, and the brightest, densest material is now oriented perpendicularly, corresponding to the slower moving equatorial material (the black contours here represent thermal emission from Day 126). This can be explained if the nova stops blowing a wind at late times, and the ejecta drift off into space. \vspace{-0.2cm} \label{v959mon}}
\end{figure}
 It is still unclear if nova explosions share a common ejecta morphology and shock production mechanism, or if their structures are shaped by a diversity of processes. Our imaging of V959\,Mon illustrates that it is difficult to understand the full complexity of nova ejecta shaping from a single snapshot (as has usually been done in the optical, often decades after explosion; \citealt{O'Brien08}), because mass loss can be prolonged over months and the ejecta exhibit a large range of densities. 
The high sensitivity and high angular resolution of the ngVLA for every observation will capture the evolution from one phase to the other in detail, and show whether our hypothesized ejecta morphology is universal. 
Very long baselines ($>$1000 km) on the ngVLA will be sensitive to synchrotron emission at higher confidence and earlier in the evolution of the nova, enabling us to directly image shocked regions and test if they are powered by the same mechanism in all novae.

\vspace{0.15in}
\noindent {\large \bf 4. Can a Diversity of Shocks Explain the Diversity of Novae?}\\
It is now well-established that $\gamma$-ray luminosity varies from nova to nova by at least two orders of magnitude \citep{Cheung16, Franckowiak18}. This is only the latest in a suite of commonly-observed but difficult-to-explain phenomena which highlight the diversity of novae. For example, in some novae the optical emission rises and declines in a matter of days, while in others the light curve bounces around maximum light for months \citep{Strope10}. Some novae have their optical emission entirely blotted out by dust forming in their ejecta, while others seem to form no dust at all \citep{Evans08}. 
This diversity remains perplexing, as theory predicts that the properties of novae should be simply determined by two properties---the white dwarf mass and the accretion rate onto it \citep{Townsley04}. Our hypothesis is that the long-standing mystery of nova diversity can be explained by complex mass ejection and shocks. This diversity presents an ideal opportunity to survey parameter space and test our models of nova shocks.

To carry out this test, we require high-quality multi-wavelength observations of a large sample of novae. Critically, \emph{Fermi}/LAT should be sustained in operation as long as possible, as its all-sky GeV $\gamma$-ray observations are ideal for constraining the diversity of shock power in novae. In addition, MeV $\gamma$-ray monitoring with the All-sky Medium Energy Gamma-Ray Observatory mission (AMEGO; \citealt{McEnery16}) will open an entirely new window to understanding which novae host strong shocks and particle acceleration. This $\gamma$-ray monitoring will be most powerful when combined with high-cadence optical/IR surveys, like ASAS-SN \citep{Shappee14}, Gattini-IR \citep{Soon18}, and LSST,  which can uncover the full nova population of the Milky Way. 

Once discovered, multi-wavelength follow-up observations are needed to measure the energetics of the ejecta and the properties of the shocks, for novae with a range of $\gamma$-ray luminosities. Fast-response and optimally-coordinated optical/IR spectroscopic and photometric observations are critical for measuring ejecta velocities and disentangling the different phases of the outburst. A state-of-the-art intelligent observatory, comprised of an army of telescopes with diverse functionality---like that planned at the South African Astronomical Observatory---would be ideal. Multi-frequency radio monitoring with the ngVLA will measure the thermal ejecta mass and reveal the energetics of lower-energy relativistic particles through synchrotron emission. X-ray monitoring with a successor to \emph{Swift} like the Transient Astrophysics Probe (TAP; \citealt{Camp16}) will provide measurements of the white dwarf and ejecta mass from super-soft X-ray emission \citep[e.g.,][]{Schwarz11}, while patrolling for thermal emission from shocked gas.

\vspace{0.15in}
\noindent {\large \bf 5. What is the Efficiency of Particle Acceleration in Dense Shocks?}\\
We expect that a small fraction of the shock power ($\lesssim$few percent; \citealt{Caprioli14}) goes into accelerating relativistic particles, while the majority remains with thermal particles. Given the expected shock velocities in novae ($\sim$1,000 km/s), thermal emission from the shock is expected to be in the form of X-rays. However, X-ray monitoring with \emph{Swift}/XRT revealed many non-detections of novae during $\gamma$-ray detection \citep{Metzger14}, implying that the shocks are behind sufficiently high absorbing columns that only X-rays above 10 keV shine through. Sensitivity to harder X-rays provided by \emph{NuSTAR} has enabled the detection of thermal emission from shocks in two novae concurrently with gamma-ray detection \citep{Nelson18, Nelson19}.

However, we are now faced with the mystery that the observed X-ray luminosity, if taken as a faithful measure of the shock power, implies an extraordinarily high efficiency of particle acceleration in nova shocks ($>$20\%; \citealt{Nelson19}). It has been proposed that the shock power may be higher than the observed X-ray luminosity because a large fraction of the shock power is emitted by lower temperature and/or more absorbed gas \citep{Steinberg18}.  This can be tested by repeated X-ray observations of a single nova with sufficient sensitivity, since the absorbing column is known to decrease with time.  We are attempting such observations, but future observatories, such as TAP, AMEGO, and the High-Energy X-ray Probe (HEX-P; \citealt{Harrison16}), should be able to provide definitive observations easily.

In an era where the entire high-energy spectrum may be observed, from keV to TeV energies (with \emph{Swift}/TAP, \emph{NuSTAR}/HEX-P, AMEGO, \emph{Fermi}, and CTA), novae present one of the best opportunities to measure the non-thermal spectrum and tie it to shock parameters. The wide-band spectrum of non-thermal emission can definitively distinguish between the hadronic or leptonic process for $\gamma$-ray emission in novae \citep{Vurm18}. Very high-energy observations with the Cherenkov Telescope Array (CTA) and the Southern Gamma-Ray Survey Observatory (SGSO) will constrain how the maximum energy of particle acceleration scales with shock velocity and density \citep{Metzger16}. There are even prospects for detecting neutrinos from novae, with an upgraded IceCube (Gen2; \citealt{Metzger16}).

\vspace{0.15in}
\noindent {\large \bf 6. Are Shocks the Primary Site of Dust Formation?}\\
A long-standing puzzle about novae concerns the observation of dust formation weeks to months after an outburst --- even though the ejecta are ionized and subject to the harsh radiation field of the central white dwarf. Dust formation requires sufficiently cool (T $\lesssim 10^3$~K) and predominantly neutral gas, in which molecules can condense to subsequently form grains.  Only a subset of novae produce dust, and among these the total mass, composition, distribution, and condensation time vary \citep{Evans08}.

If dense radiative shocks are commonplace in novae, however, they could be an ideal environment for dust formation. Gas behind the shock cools efficiently and the density increases accordingly, leading to the formation of cold, dense, neutral clumps of gas that are protected from the harsh radiation from the white dwarf, allowing rapid formation of grains that grow efficiently to large sizes \citep{Derdzinski17}.  Multi-wavelength observations that concurrently observe shock-powered emission (in $\gamma$-rays, hard X-rays, and non-thermal radio emission; \S2--4) and dust signatures (in the IR)
can test the relationship between shock formation and dust condensation.  Such observations can also directly constrain the density and temperature of post-shock gas, further revealing whether the environment in which shocks are formed could be conducive to dust production.  Moreover, time-domain infrared surveys like Gattini-IR \citep{Soon18} have the potential to provide unprecedented statistics on the fraction of novae that produce dust and the fraction that do so along our line of sight, constraining the dust-covering fraction and ejecta morphology.  A deep understanding the relationship between shocks and dust production -- as can be gleaned through investigations of shocks in novae -- is also crucial for the study of other dusty and obscured transients (such as the so-called SPRITES; \citealt{Kasliwal17}) as we
continue through the era of time-domain astrophysics.

\vspace{0.15in}
\noindent {\large \bf 7. Conclusion: Novae as High-Energy Laboratories}\\
Novae expand our understanding of astrophysical shocks to high densities and low velocities, and are powerful laboratories for understanding other potentially shock-powered transients including Type~IIn and super-luminous supernovae \citep[e.g.][]{2014ApJ...788..154O,2018SSRv..214...27C}, stellar mergers \citep{Metzger17, 2018ApJ...858...53L}, and exotic dusty  transients \citep{Kasliwal17}. Novae are nearby and luminous enough that they can be imaged with high spatial resolution and monitored across the entire electromagnetic spectrum from radio to $\gamma$-rays. With these high-quality observations, we will be able to understand the conditions and EM signatures of radiative shocks, dust formation, and particle acceleration in real time and to unprecedented precision.


\begin{multicols}{2}
  \setlength{\bibsep}{0pt plus 0.3ex}
  \bibliographystyle{mnras_shortlisthack}
  \bibliography{ChomiukLaura}
\end{multicols}

\end{document}